\documentclass[%
  10pt,
  superscriptaddress,
  twocolumn,
  bibnotes,
  amsfonts,
  amssymb,
  amsmath,
  aps,
  prl,
  floatfix
]{revtex4-2}
\usepackage{mathtools}
\usepackage{mathrsfs}
\usepackage{amsthm}
\usepackage{bm}
\usepackage{braket}

\usepackage{graphicx}
\usepackage{wrapfig}
\usepackage{tikz}
\usetikzlibrary{quantikz2}

\usepackage{array}
\usepackage{dcolumn}
\usepackage{multirow}
\usepackage{booktabs}
\usepackage[inline]{enumitem}

\usepackage{csquotes}
\usepackage{comment}
\usepackage{xcolor}
\usepackage[normalem]{ulem}
\usepackage{orcidlink}
\usepackage{pdfpages}
\makeatletter
\AtBeginDocument{\let\LS@rot\@undefined}
\makeatother

\usepackage{xurl}
\usepackage{hyperref}
\hypersetup{
  colorlinks=true,
  citecolor=cyan,
  urlcolor=cyan,
  linkcolor=cyan,
  breaklinks=true,
}

\newtheoremstyle{break}
  {} {} {\itshape} {} {\bfseries} {.} {\newline} {\thmname{#1}\thmnumber{ #2}\thmnote{ (\bfseries #3)}}
\theoremstyle{break}

\newtheorem{theorem}{Theorem}

\DeclareMathOperator{\Tr}{Tr}

\let\oldd\d \renewcommand{\d}{\ifmmode\mathrm{d}\else\oldd\fi}
\let\oldi\i \renewcommand{\i}{\ifmmode\mathrm{i}\else\oldi\fi}

\newcommand{\expt}[1]{\langle #1 \rangle}
\newcommand{\ketbra}[2]{| {#1} \rangle\langle {#2} |}

\newcommand{\ti}[1]{\textit{#1}}

\newcommand{\ra}{\rightarrow}

\newcommand{\eps}{\epsilon}
\newcommand{\veps}{\varepsilon}

\newcommand{\md}{\textemdash}
\newcommand{\nd}{\textendash}

\newcommand{\eq}[1]{\begin{align} #1 \end{align}}
\newcommand{\subeq}[1]{\begin{subequations} #1 \end{subequations}}

\newcommand{\bbR}{\mathbb{R}}

\makeatletter
\@tfor\Letter:=ABCDEFGHIJKLMNOPQRSTUVWXYZ\do{%
  \expandafter\edef\csname rm\Letter\endcsname{\noexpand\mathrm{\Letter}}
  \expandafter\edef\csname bf\Letter\endcsname{\noexpand\mathbf{\Letter}}
  \expandafter\edef\csname sf\Letter\endcsname{\noexpand\mathsf{\Letter}}
  \expandafter\edef\csname cal\Letter\endcsname{\noexpand\mathcal{\Letter}}
  \expandafter\edef\csname scr\Letter\endcsname{\noexpand\mathscr{\Letter}}
}
\makeatother

{\list{}{\leftmargin=#1\rightmargin=#1}\item[]}%
{\endlist}

\definecolor{lightblue}{rgb}{0.678, 0.847, 0.902}
\definecolor{lightgreen}{rgb}{0.565, 0.933, 0.565}
\definecolor{lightyellow}{rgb}{1.000, 1.000, 0.600}
\definecolor{lightpurple}{rgb}{0.867, 0.627, 0.867}
\definecolor{lightorange}{rgb}{1.000, 0.753, 0.502}
\definecolor{lightpink}{rgb}{1.000, 0.714, 0.757}
\definecolor{lightred}{rgb}{1.000, 0.714, 0.757}
\definecolor{lightcyan}{rgb}{0.878, 1.000, 1.000}
\definecolor{lightcoral}{rgb}{0.941, 0.502, 0.502}
\definecolor{lightsalmon}{rgb}{1.000, 0.627, 0.478}

\begin{document}

\title{Error and Disturbance as Irreversibility with Applications:\protect\\Unified Definition, Wigner\nd Araki\nd Yanase Theorem and Out-of-Time-Order Correlator}

\author{
Haruki Emori
\orcidlink{0009-0007-2264-9192}
}
\email{emori.haruki.i8@elms.hokudai.ac.jp}
\affiliation{
Graduate School of Information Science and Technology, Hokkaido University, Kita 14, Nishi 9, Kita-ku, Sapporo, Hokkaido 060-0814, Japan
}
\affiliation{
RIKEN Center for Interdisciplinary Theoretical and Mathematical Sciences (iTHEMS), RIKEN, 2-1 Hirosawa, Wako, Saitama, 351-0198 Japan
}
\author{
Hiroyasu Tajima
}
\email{hiroyasu.tajima@uec.ac.jp}
\affiliation{
Department of Informatics, Faculty of Information Science and Electrical Engineering, Kyushu University, 744 Motooka, Nishi-ku, Fukuoka 819-0395, Japan
}
\affiliation{
JST, FOREST, 4-1-8, Honcho, Kawaguchi, Saitama 332-0012, Japan
}
\date{\today}

\begin{abstract}
Defining an error of measurement has long been a foundational problem in science: even in classical experiments, data are statistical and admit no single universally optimal definition of error.
In quantum mechanics, the challenge deepens: observed entities often lack preexisting definite values, and the act of measurement unavoidably disturbs the system of interest. Consequently, both error and disturbance must be quantified, and various definitions have been proposed to date.
However, a unified perspective for understanding the differences and similarities among these diverse definitions of error and disturbance, and an operational framework for distinguishing between them, remain elusive.
In this Letter, we propose a novel framework for defining error and disturbance using irreversibility.
Our framework converts the error and disturbance of a quantum measurement of a system under consideration into the irreversibility of an ancillary qubit system, using a quantum comb composed of loss and recovery processes.
The mechanism enables us to make the operational distinction that error uses the classical outputs, while disturbance uses the quantum outputs of the measurement in the recovery process.
Furthermore, our framework yields several key consequences: (i) it encompasses existing definitions, (ii) it establishes a universal constraint on error and disturbance defined by any measure of an arbitrary quantum process under a conservation law, and (iii) it reveals an operational connection between irreversibility and the out-of-time-ordered correlator (OTOC), a metric of quantum chaos.
It also provides a constraint on the OTOC under a conservation law and a method for its experimental evaluation, which is demonstrated on a quantum processor.
\end{abstract}

\maketitle

\textit{\textbf{Introduction.---}}
Defining an error of measurement has long been recognized as a fundamental problem in science.
Even before the advent of quantum mechanics, data obtained in real experiments come with statistical uncertainty, and---as Gauss already observed---while there is not a priori correct judgment about which measure of reliability is most effective for such statistical data, there is a utility of establishing a standard measure \cite{Gauss12}.

In the quantum regime, the problem becomes even more complex. 
A measurement of an observable in a state, which is not an eigenstate of it, involves an inherently statistical treatment of outcomes.
Moreover, the very act of measurement unavoidably alters a system of interest, as Heisenberg emphasized in formulating the uncertainty principle.
It brings about a need for quantifying not only error but also disturbance.

Because of their importance, to date, error and disturbance have been framed by various approaches: Arthurs--Kelly--Goodman (AKG) \cite{Arthurs65,Yamamoto86,Arthurs88,Ishikawa91,Ozawa91}, Ozawa \cite{Ozawa03,Ozawa04,Ozawa19,Ozawa21}, Watanabe--Sagawa--Ueda (WSU) \cite{Watanabe11PRA,Watanabe11ARX}, Busch--Lahti--Werner (BLW) \cite{Busch13,Busch14PRA,Busch14RMP}, and Lee--Tsutsui (LT) \cite{Lee20ARX,Lee20ENT,Lee22}.
While these efforts have been successful within their specific frameworks, they make it difficult to understand the fundamental similarities and differences between error and disturbance across different definitions.
A single, operationally clear way to distinguish between them has been demanded.

In this Letter, we propose a novel framework that leverages irreversibility, which is another fundamental notion in physics, to address this conundrum.
Our approach introduces a qualitatively new capability: the ability to clearly and operationally distinguish between error and disturbance.
We achieve this by applying a channel conversion---a quantum comb \cite{Chiribella08}---to the measuring process, effectively converting the error and disturbance in a target system into the irreversibility in an ancillary qubit system through a loss and recovery process.
With this mechanism, our framework encompasses existing frameworks through a common language, irreversibility, and distinguishes each of them by their differing loss and recovery processes.
We can say that error is irreversibility revealed when recovery is attempted using only the classical outputs of the measurement.
In other words, error is the difference between the value being measured and the value actually measured.
On the other hand, disturbance is irreversibility revealed when recovery is attempted using only the quantum outputs of the measurement.
In other words, disturbance is the difference between the operator being affected and the operator actually affected.
This distinction not only provides a comprehensive view of existing definitions but also clarifies their essence in a way that has not been possible before.

Furthermore, our framework yields several important consequences.
(i) we unify the existing frameworks: we can derive that of AKG, Ozawa, WSU, BLW, and LT from ours.
(ii) we extend the Wigner--Araki--Yanase (WAY) theorem \cite{Wigner52,Araki60,Ozawa02m,Korzekwa13,Marvian12,Ahmadi13NJP,Tajima19,Tajima22,Kuramochi23,Mohammady23,Tajima25,Tajima25ARX}, a universal restriction on measurement under a conservation law, to a quantitative one for arbitrary definitions of error and disturbance and arbitrary quantum processes.
While previous theorems have been limited to the Ozawa-type error \cite{Ozawa02m,Korzekwa13,Tajima19} and the gate fidelity error \cite{Tajima22}, we solved the open problem of extending it.
(iii) we establish a fundamental link between the irreversibility and the out-of-time-ordered correlator (OTOC) \cite{Larkin69}, which is a measure of quantum chaos in high-energy and condensed matter physics \cite{Maldacena16}.
The link allows us to provide a general bound on the OTOC when the scrambling dynamics obey a conservation law.

Our work paves the way for a direct application of decades of knowledge from non-equilibrium physics to foundations of quantum physics and high-energy physics.
In particular, we propose a simple experimental method for evaluating the OTOC that only requires sampling at the end of the protocol.
In a companion paper, we report a proof-of-principle experiment on a quantum computer that demonstrates the effectiveness of our method \cite{Emori25}.

\ti{\textbf{Our formulation.\md}}In order to formulate error and disturbance as irreversibility, we construct irreversibility evaluation protocols (IEPs) for quantum processes in finite dimensional systems.
Therefore, we begin by introducing the measure of irreversibility.

\ti{Irreversibility of quantum processes:}
So far, several methods of quantifying the irreversibility have been developed.
Above all, the measure $\delta$ of irreversibility in Ref.~\cite{Tajima22} has the benefit that it encompasses a variety of irreversibility measures.
As an instance, the measure $\delta$ gives a minimum threshold for the entropy production \cite{Funo18} in stochastic thermodynamics, a lower bound for the entanglement fidelity error \cite{Watrous18} in quantum error correction, and a recovery error of the Petz recovery map \cite{Hayden04,Wilde15,Junge18}.
For this reason, we adopt the measure $\delta$ defined by Ref.~\cite{Tajima22}.

Let us consider a quantum process described by a completely positive trace-preserving (CPTP) map $\calL$ from a target system $\bfS$ to another system $\mathbf{S'}$ and an arbitrary test ensemble $\Omega=\{p_{k},\rho_{k}\}$, where $\{\rho_{k}\}$ is a set of quantum states in $\bfS$ with a set $\{p_{k}\}$ of preparation probabilities.
Then, we define the irreversibility of $\calL$ with respect to $\Omega$ as follows \cite{Tajima22}:
\eq{
\label{eq:irreversibility}
\delta(\calL,\Omega)&:=\min_{\calR:\mathbf{S'}\ra \bfS}\sqrt{\sum_{k}p_{k}\delta^{2}_{k}},\\
\label{eq:distance}
\delta_{k}&:=D_{F}(\rho_{k},\calR\circ\calL(\rho_{k})).
}
Here, the minimization is performed over CPTP maps $\calR$ from $\mathbf{S'}$ to $\bfS$, $D_{F}(\rho,\sigma):=\sqrt{1-F(\rho,\sigma)^{2}}$ is the purified distance \cite{Tomamichel12}, and $F(\rho,\sigma):=\Tr(\sqrt{\sqrt{\rho}\sigma\sqrt{\rho}})$ is the Uhlmann fidelity \cite{Uhlmann76}.
Focusing on a specific CPTP map $\calR'$, the irreversibility of $\calL$ regarding $\Omega$ is
\eq{
\label{eq:specific_irreversibility}
\delta(\calL,\calR',\Omega)=\sqrt{\sum_{k}p_{k}\delta^{2}_{k}}.
}
It is important to note that $\delta(\calL,\Omega)\le\delta(\calL,\calR',\Omega)$ always holds, by definition.
Under these settings, we proceed to present our main result.

\begin{figure}[tb]
\centering
\includegraphics[width=.9\linewidth]{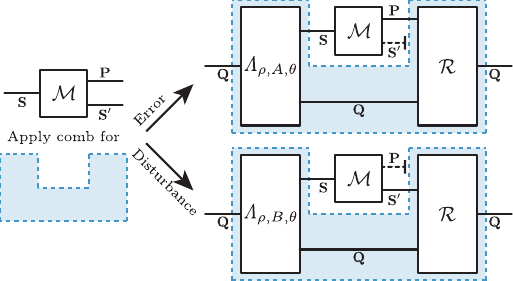}
\caption{
A schematic of irreversibility evaluation protocols for general definition of error and disturbance, using quantum combs.
Each protocol consists of loss process, $\calL_{\rho,A,\theta,\calP}:=\calP_{\calM}\circ\mathit{\Lambda}_{\rho,A,\theta}$ for error and $\calL_{\rho,B,\theta,\calI}:=\calI_{\calM}\circ\mathit{\Lambda}_{\rho,B,\theta}$ for disturbance, and recovery process $\calR$.
The measuring processes $\calM$ are distinguished between two types, $\calP_{\calM}$ and $\calI_{\calM}$, depending on the outputs\md either classical outcomes in $\bfP$ or quantum states in $\mathbf{S'}$.
}
\label{fig:Quantum_comb_gen_def}
\end{figure}

\ti{General definition of error and disturbance as irreversibility:}
Similar to how quantities related to irreversibility are treated in a unified manner by Eq.~\eqref{eq:irreversibility}, we demonstrate that various definitions of error and disturbance can also be treated by Eq.~\eqref{eq:irreversibility}.
The main idea of our formulation is to design the IEPs based on a framework proposed for direct evaluations of Ozawa's error \cite{Hofmann21} and disturbance \cite{Emori24}.
The framework introduces an ancillary qubit system $\bfQ$ in addition to the target system $\bfS$ with finite dimension.
By applying weak interactions before and after the measurement, the information of the measurement in $\bfS$ is transcribed into the information of the time evolution in $\bfQ$.
In analogy to this, by inserting additional processes to and reshaping the measuring process, as a \ti{comb} (Fig.~\ref{fig:Quantum_comb_gen_def}), the IEPs establish the universal mechanism of that transcription.
The mechanism allows us to evaluate the error and disturbance of the measurement in $\bfS$ by looking at the irreversibilities (more precisely, these derivatives) of the loss process in $\bfQ$.
The IEPs consist of two components:
\begin{enumerate*}[label=(P\arabic*)]
    \item a loss process $\calL$, including the weak interaction on $\bfS+\bfQ$ and the subsequent measurement on $\bfS$, to make $\bfS+\bfQ$ have a correlation and cause a loss to $\bfQ$; and
    \item a recovery process $\calR$, including the inverse mapping that exploits information obtained by the measurement, to restore $\bfQ$ up to its original state.
\end{enumerate*}
Hence, the IEPs make a one-to-one correspondence between an uncertainty of the information gained by the measurement and an incompleteness of the recovery.

Suppose a measurement $\calM$ of an observable $A$ in $\bfS$ with an initial state $\rho$ and its process is described by measurement operators $\{M_{m}\}$ mapping from $\bfS$ to $\mathbf{S'}$, where $\{m\}$ are measurement outcomes; and denote $B$ as another observable.
With these notations, we define the error $\veps$ and disturbance $\eta$ of $\calM$ by
\subeq{
\label{eq:general_def}
\eq{
\label{eq:general_error}
\veps^{2}(\rho,A,\calM)&:=\lim_{\theta\ra0}\frac{\delta^{2}(\calL_{\rho,A,\theta,\calP},\Omega_{1/2,\pm})}{\theta^{2}},\\
\label{eq:general_dist}
\eta^{2}(\rho,B,\calM)&:=\lim_{\theta\ra0}\frac{\delta^{2}(\calL_{\rho,B,\theta,\calI},\Omega_{1/2,\pm})}{\theta^{2}},
}
}
where $\Omega_{1/2,\pm}$ is a specific test ensemble $\{(1/2,1/2),(\ket{+},\ket{-})\}$, $\ket{\pm}$ are eigenstates of the Pauli-$x$ operator $\sigma_{x}$ in $\bfQ$, and
\subeq{
\label{eq:loss_proc}
\eq{
\label{eq:loss_proc_error}
\calL_{\rho,A,\theta,\calP}&:=\calP_{\calM}\circ\mathit{\Lambda}_{\rho,A,\theta}=\calP_{\calM}\circ\calU_{A,\theta}\circ\calA_{\rho},\\
\label{eq:loss_proc_dist}
\calL_{\rho,B,\theta,\calI}&:=\calI_{\calM}\circ\mathit{\Lambda}_{\rho,B,\theta}=\calI_{\calM}\circ\calU_{B,\theta}\circ\calA_{\rho}
}
}
are composite processes from $\bfQ$ to $\bfP+\bfQ$ and from $\bfQ$ to $\bfS'+\bfQ$, respectively.
$\bfP$ is a memory system for outcomes of $\calM$.
Specifically, the loss process $\calL_{\rho,A,\theta,\calP}$ is implemented by
\begin{enumerate*}[label=(E\arabic*)]
    \item an appending process $\calA_{\rho}(\bullet):=\rho\otimes\bullet$ which adds the state $\rho$ of $\bfS$ to $\bfQ$;
    \item an unitary process $\calU_{A,\theta}$ on $\bfS+\bfQ$, where $\calU_{X,\theta}(\bullet):=e^{-i\theta X\otimes\sigma_{z}}(\bullet)e^{i\theta X\otimes\sigma_{z}}$ generated by an arbitrary observable $X$ and the Pauli-$z$ operator $\sigma_{z}$; and
    \item a measuring process $\calP_{\calM}(\bullet):=\sum_{m}\Tr[M_{m}(\bullet)M^{\dagger}_{m}]\ketbra{m}{m}_{\bfP}$ from $\bfS$ to $\bfP$.
\end{enumerate*}
Similarly, the other loss process $\calL_{\rho,B,\theta,\calI}$ is implemented by
\begin{enumerate*}[label=(D\arabic*)]
    \item an appending process $\calA_{\rho}$;
    \item an unitary process $\calU_{B,\theta}$; and
    \item a measuring process $\calI_{\calM}(\bullet):=\sum_{m}M_{m}(\bullet)M^{\dagger}_{m}$ from $\bfS$ to $\mathbf{S'}$, where $\calI_{\calM}$ corresponds to the quantum instrument \cite{Davies70,Ozawa84}.
\end{enumerate*}

We remark that $\calP_{\calM}$ and $\calI_{\calM}$ can be obtained by taking partial traces of $\mathbf{S'}$ and $\bfP$, respectively, from the output of $\calM$.
The difference between $\calL_{\rho,A,\theta,\calP}$ and $\calL_{\rho,B,\theta,\calI}$ arises from a way of mathematical description of the measuring process, i.e. what is chosen as the output, either outcomes or post-measurement states, of $\calM$.

According to Eq.~\eqref{eq:specific_irreversibility}, the error and disturbance for specific recovery processes $\calR'$ are naturally of the forms
\subeq{
\label{eq:specific_def}
\eq{
\label{eq:specific_error}
\veps^{2}(\rho,A,\calM,\calR')&:=\lim_{\theta\ra0}\frac{\delta^{2}(\calL_{\rho,A,\theta,\calP},\calR',\Omega_{1/2,\pm})}{\theta^{2}},\\
\label{eq:specific_dist}
\eta^{2}(\rho,B,\calM,\calR')&:=\lim_{\theta\ra0}\frac{\delta^{2}(\calL_{\rho,B,\theta,\calI},\calR',\Omega_{1/2,\pm})}{\theta^{2}}.
}
}
By virtue of our formulation, we highlight three applications below.

\ti{Application 1. The unification and derivation of the existing formulations for error and disturbance:}
\begin{table}[tb]
\caption{\label{tab:Applications_1_and_3}
A list of existing formulations of error, disturbance, and OTOC that can be derived from our formulation.
}
\begin{ruledtabular}
\begin{tabular}{llll}
Defs. & $\calL$ \& $\calR$ & Conditions & Relationships \\
\colrule
AKG & $\calR_{M,\bfP/B,\bfS}$ & Unbiasedness & $\subset$ Ozawa \& LT \\
Ozawa & $\calR_{M,\bfP/B,\bfS}$ & 
\begin{tabular}{l}
$M=\sum_{m}f(m)\ketbra{m}{m}$\\s.t. $f(m)=A(m)$
\end{tabular}
& \\
WSU & $\min_{X}$,$\calR_{X}$ & Local-representability & $\subset$ LT \\
BLW & $\calR_{M,\bfP/B,\bfS}$ & Calibration \& $\rho^{\otimes2}$ & \\
LT & $\min_{X}$,$\calR_{X}$ & Real-valued outcomes & $\leftrightarrow$ Ozawa \\
OTOC & $\calD_{W}$ \& $\calR_{V,\bfS}$ & & \\
\end{tabular}
\end{ruledtabular}
\end{table}
From the perspectives of state-dependence (or not), physical significance and experimental feasibility, the formulations of error and disturbance have been proposed by Arthurs--Kelly--Goodman (AKG) \cite{Arthurs65,Yamamoto86,Arthurs88,Ishikawa91,Ozawa91}, Ozawa \cite{Ozawa03,Ozawa04,Ozawa19,Ozawa21}, Watanabe--Sagawa--Ueda (WSU) \cite{Watanabe11PRA,Watanabe11ARX}, Busch--Lahti--Werner (BLW) \cite{Busch13,Busch14PRA,Busch14RMP}, and Lee--Tsutsui (LT) \cite{Lee20ARX,Lee20ENT,Lee22}.
Nevertheless, a formulation unifying all these approaches has not been known.
In the form of solving this problem, our formulation subsumes the existing formulations, as summarized in Table.~\ref{tab:Applications_1_and_3}.
Note that we here assume the calibration measure for BLW and the real-valued outcomes for LT, which are handled in standard experiments; the cases where these constraints are relaxed remain unresolved problems.

\begin{figure}[tb]
\centering
\includegraphics[width=.8\linewidth]{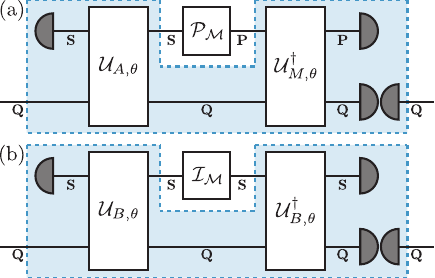}
\caption{Irreversibility evaluation protocols for Ozawa's (a) error and (b) disturbance.
}
\label{fig:Quantum_comb_ozawa_def}
\end{figure}
To reproduce the existing formulations from ours, we introduce a crucial recovery process $\calR_{X,\bullet}:=\calJ_{\bullet}\circ\calU^{\dagger}_{X,\theta}$, a CPTP map from $\mathbf{\bullet}+\bfQ$ to $\bfQ$, where $\calJ_{\bullet}(\cdot):=\sum_{j=\pm}\bra{j}\Tr_{\bullet}[(\cdot)]\ket{j}\ketbra{j}{j}_{\bfQ}$ for $\bullet=\bfP,\mathbf{S'}$.
Using $\calR_{X,\bullet}$, aforementioned errors and disturbances are strictly covered by $\veps(\rho,A,\calM,\calR_{X,\bfP})$ and $\eta(\rho,B,\calM,\calR_{X',\mathbf{S'}})$.
Additionally, the bounds
\subeq{
\eq{
\veps(\rho,A,\calM)&\le\veps(\rho,A,\calM,\calR_{X,\bfP}),\\
\eta(\rho,B,\calM)&\le\eta(\rho,B,\calM,\calR_{X',\mathbf{S'}})
}
}
hold for any $X$ and $X'$.
Since Ozawa's error and disturbance subsume others, we focus on demonstrating the derivation of Ozawa's one to facilitate intuitive understanding.
The case of other forms and further expositions can be found in the Supplemental Material \footnote{See Supplemental Material for detailed proofs, extended discussions, and additional information.}\setcounter{footnote}{1}.

\ti{Example. Ozawa’s error and disturbance:}
Consider a measurement $\{M_{m}\}$ of $A$ in $\bfS$ and a set of real-valued functions $\{A(m)\}$ based on the measurement outcomes $\{m\}$.
Then, Ozawa's error and disturbance are defined by
\subeq{
\label{eq:Ozawa_def}
\eq{
\label{eq:Ozawa_error}
\veps_{\text{O}}(A)&:=\sum_{m}\|M_{m}[A-A(m)]\sqrt{\rho}\|_{2},\\
\label{eq:Ozawa_dist}
\eta_{\text{O}}(B)&:=\sum_{m}\|[M_{m},B]\sqrt{\rho}\|_{2};
}
}
where $\bfS=\mathbf{S'}$ \cite{Ozawa03,Ozawa04,Ozawa05JOB}.
If we apply $\calR_{M,\bfP}$ to Eq.~\eqref{eq:specific_error} with $M=\sum_{m}f(m)\ketbra{m}{m}$ such that $f(m)=A(m)$, and $\calR_{B,\bfS}$ to Eq.~\eqref{eq:specific_dist}, we obtain the equalities
\subeq{
\label{eq:Ozawa_def=our_def}
\eq{
\label{eq:Our_error=Ozawa_error}
\veps(\rho,A,\calM,\calR_{M,\bfP})&=\veps_{\text{O}}(A),\\
\label{eq:Our_dist=Ozawa_dist}
\eta(\rho,B,\calM,\calR_{B,\bfS})&=\eta_{\text{O}}(B)
}
}
in addition to the obvious inequalities $\veps(\rho,A,\calM)\le\veps_{\text{O}}(A)$ and $\eta(\rho,B,\calM)\le\eta_{\text{O}}(B)$.
The procedures are illustrated in Fig.~\ref{fig:Quantum_comb_ozawa_def}(a) and (b).
Applying $\calU^{\dagger}_{M,\theta}$ becomes possible to restore the relative phase shift of the state in $\bfQ$ caused by $\calL_{\rho,A,\theta,\calP}$, proportional to the amount $f(m)$.
By contrast, applying $\calU^{\dagger}_{B,\theta}$ enables us to check the commutativity between $M_{m}$ and the generator $B$ of $\calU_{B,\theta}$.
If these two operators commute, it means that we can perfectly restore the relative phase shift by $\calL_{\rho,B,\theta,\calI}$.

\ti{Application 2. The extended WAY theorem for errors and disturbances of arbitrary definitions and processes:}
With the help of the versatility of our formulation, we solve a long-standing open problem\md an extension of the WAY theorem to errors and disturbances of arbitrary definitions and processes, as sketched in Table.~\ref{tab:Applications_2}.
The WAY theorems \cite{Tajima22,Ozawa02m,Korzekwa13,Marvian12,Ahmadi13NJP,Tajima19,Kuramochi23,Mohammady23} state, in the presence of an additive conservation law, implementations of a projective measurement for an observable that does not commute with the conserved one are impossible.
Although the quantitative WAY theorems for finite error have been actively studied \cite{Tajima22,Ozawa02m,Korzekwa13,Tajima19}, these theorems for quantum measurements are limited to the Ozawa-type error \cite{Ozawa02m,Korzekwa13,Tajima19} and gate-fidelity error \cite{Tajima22}; the quantitative WAY theorem for error and disturbance of arbitrary definitions has thus remained an enormous challenge \footnote{More precisely, the quantitative WAY theorems \cite{Tajima22,Ozawa02m,Korzekwa13,Tajima19} are given for the error defined as the expectation value of square of the ``error operator''. But this error is easily shown to be equivalent to Ozawa's error \cite{Note1}.}\setcounter{footnote}{2}.
By combining our definition with the symmetry\nd irreversibility\nd quantum coherence (SIQ) trade-off relation \cite{Tajima22}, we establish a theorem overcoming the restriction\md errors defined by a limited scope and disturbances hitherto entirely unknown.
\begin{table}[tb]
\caption{\label{tab:Applications_2}
Extensions of the WAY theorem\footnote{The arrow means an extension of the results from ``before our formulation'' to ``after our formulation''
}.
}
\begin{ruledtabular}
\begin{tabular}{lcc}
Processes & Defs. of error & Defs. of disturbance \\
\colrule
\vspace{-2ex}\\
Arbitrary &
Not exist\footnote{When we consider an arbitrary process, the WAY-type tradeoff relations between error and resource do not exist (There is a counter example).} &
\multirow{3}{*}{\begin{tikzpicture}
\node[draw, dashed, rectangle, inner xsep=3pt, inner ysep=3pt, align=left] (m) {
Unknown \\[1.6ex]
Unknown \\[3.6ex]
Unknown
};
\draw[dashed, ->] (m.south) -- ++(0,-0.2) node[anchor=north] {};
\end{tikzpicture}} \\
Measurement & \multirow{2}{*}{\begin{tikzpicture}
\node[draw, dashed, rectangle, inner xsep=3pt, inner ysep=3pt, align=left] (m) {
Ozawa-type \& \\[0.5ex]
Gate fidelity-type \\[0.5ex]
Gate fidelity-type
};
\draw[dashed, ->] (m.south) -- ++(0,-0.2) node[anchor=north] {};
\end{tikzpicture}} & \\[2.5ex]
Unitary & & \\[2.5ex]
& \begin{tabular}{c}
Arbitrary
\end{tabular} & \begin{tabular}{c}
Arbitrary
\end{tabular} \\
\end{tabular}
\end{ruledtabular}
\end{table}

To show our result, we introduce the SLD-quantum Fisher information for the state family $\{e^{-i\eps X}\rho e^{i\eps X}\}_{\eps\in\bbR}$, which is a measure of quantum coherence in the resource theory of asymmetry \cite{Hansen08,Marvian13,Zhang17,Takagi19,Marvian20,Yamaguchi23Q,Marvian22,Yamaguchi23PRL,Kudo23,Shitara23,Yamaguchi24,Yamashika25,kazi25}:
\eq{
\calF_{\rho}(X)=4\lim_{\eps\ra0}\frac{D^{2}_{F}(e^{-i\eps X}\rho e^{i\eps X},\rho)}{\eps^{2}}.\nonumber
}
The quantity indicates the amount of quantum fluctuation of the observable $X$ in the state $\rho$ \cite{Luo05,Yu13,Marvian20,Marvian22,Kudo23}.

\begin{theorem}\label{th:WAY}
Consider a measurement $\calM$ of an observable $A$ in $\bfS$.
We realize the measuring process by $\calP_{\calM}$ for error- and $\calI_{\calM}$ for disturbance-evaluation under the conservation law of the conserved charges $X_{\bullet}$ $(\bullet=\bfS,\mathbf{S'},\bfP)$. 
Then, the following inequalities hold for any state $\rho$ on $\bfS$:
\subeq{
\label{eq:WAY}
\eq{
\label{eq:WAY_error}
\veps(\rho,A,\calM)&\ge\frac{|\expt{[Y_{\bfS},A]}_{\rho}|}{\sqrt{\calF^{\text{cost}}_{\calP_\calM}}+\Delta_{F}},\\
\label{eq:WAY_dist}
\eta(\rho,B,\calM)&\ge\frac{|\expt{[Y'_{\bfS},B]}_{\rho}|}{\sqrt{\calF^{\text{cost}}_{\calI_{\calM}}}+\Delta'_{F}}.
}
}
Here, $\calF^{\text{cost}}_{\calN}$ is the implementation resource cost of a CPTP map $\calN$ from $\bm{\alpha}$ to $\bm{\alpha'}$ under the conservation law of $X_{\bullet}$ for $\bullet=\bm{\alpha},\bm{\alpha'},\bm{\beta},\bm{\beta'}$ and defined as
\eq{
\calF^{\text{cost}}_{\calN}:=\min\{\calF_{\rho_{\bm{\beta}}}(X_{\bm{\beta}})|(\rho_{\bm{\beta}},X_{\bm{\beta}},X_{\bm{\beta'}},U)\ra\calN\},
}
where the quadruple $(\rho_{\bm{\beta}},X_{\bm{\beta}},X_{\bm{\beta'}},U)$ runs implementations of $\calN$ via $\calN(\bullet)=\Tr_{\bm{\beta'}}[U(\bullet\otimes\rho_{\bm{\beta}})U^{\dagger}]$ and satisfies the conservation law
\eq{
U^\dagger(X_{\bm{\alpha'}}+X_{\bm{\beta'}})U=X_{\bm{\alpha}}+X_{\bm{\beta}}.
}
The quantities $Y_{\bfS}$, $Y'_{\bfS}$, $\Delta_{F}$, and $\Delta'_{F}$ are defined as $Y_{\bfS}:=X_{\bfS}-\calP^{\dagger}_{\calM}(X_{\bfP})$, $Y'_{\bfS}:=X_{\bfS}-\calI^{\dagger}_{\calM}(X_{\mathbf{S'}})$, $\Delta_{F}:=\sqrt{\calF_{\rho}(X_{\bfS})}+2\sqrt{V_{\calP_{\calM}(\rho)}(X_{\bfP})}$ and  $\Delta'_{F}:=\sqrt{\calF_{\rho}(X_{\bfS})}+2\sqrt{V_{\calI_{\calM}(\rho)}(X_{\mathbf{S'}})}$, where $V_\sigma(Z)$ is the variance of $Z$ in $\sigma$.
\end{theorem}
We should emphasize that Theorem~\ref{th:WAY} does not assume the Yanase condition introduced in Ref.~\cite{Ozawa02m} and the contribution of our formulation is not limited to the WAY theorem for measurements, but for other arbitrary processes.
For arbitrary processes, there are no error-cost trade-off relations (there are counterexamples), except for unitary processes \cite{Ozawa02c,Tajima18,Tajima20,Tajima21,Tajima22} and its variants \cite{Tajima22}.
Even so, the existence of disturbance-cost trade-off relations for arbitrary processes has not been denied; and now we obtain them as corollaries of Eq.~\eqref{eq:WAY_dist} by noting that any CPTP map can be described as $\calI_{\calM}$.

Although Eq.~\eqref{eq:WAY_error} provides an error-cost trade-off relation for measurements without the Yanase condition, when we assume $X_{\bfP}$ satisfies $[X_{\bfP},\ketbra{m}{m}_{\bfP}]=0$ for any $m$ where $\{\ket{m}_{\bfP}\}$ are given in the definition of $\calP_{\calM}$, in other words, when we make an assumption corresponding to the Yanase condition, we can simplify and tighten Eq.~\eqref{eq:WAY_error} as
\eq{
\label{eq:WAY_error_imroved}
\veps(\rho,A,\calM)&\ge\frac{|\expt{[X_{\bfS},A]}_{\rho}|}{\sqrt{\calF^{\text{cost}}_{\calP_{\calM}}+\calF_{\rho}(X_{\bfS})}}.
}
The elaborated proof of Theorem~\ref{th:WAY} and Eq.~\eqref{eq:WAY_error_imroved} is provided in the Supplemental Material \cite{Note1}.

\ti{Application 3. The novel treatment of OTOC as irreversibility and its experimental evaluation method:}
From the viewpoint of information propagation by interactions, we find that the OTOC is linked to the disturbance.
The OTOC quantifies the degree of information scrambling \cite{Hayden07,Sekino08,Shenker14,Shenker15} in connection with quantum chaos based on a notion of operator spreading \cite{Aleiner16,Roberts17,Keyserlingk18,Nahum18,Rakovszky18} in situations where a local observable $W$ becomes correlated with a distant one $V$, by a global interaction in dynamical quantum systems.
On the other hand, the disturbance quantitatively expresses how much the observable $B$ is affected by the measuring interaction of the measurement of the observable $A$.
In this context, there exist a well-established theory\md quantum perfect correlations \cite{Ozawa05PLA,Ozawa06}\md for evaluating the correlation between observables or probability operator-valued measures (POVMs). 
In these regards, it can be seen that they share the spirit of the quantification of observable correlations.
Take account of this aspect, we explicitly associate the OTOC with the disturbance and uniformly describe them in our formulation by providing a specific IEP.

\begin{figure}[tb]
\centering
\includegraphics[width=\linewidth]{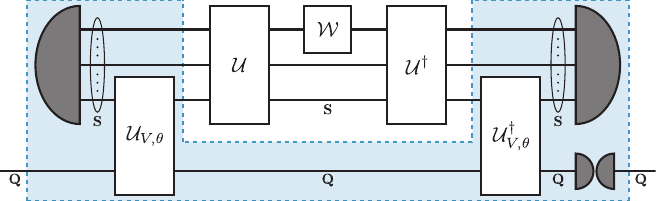}
\caption{
Irreversibility evaluation protocol for OTOC.
}
\label{fig:Quantum_comb_otoc}
\end{figure}

Let us denote $W(0)$ and $V(0)$ are (local) observables at $t=0$ supported on a small number of sites, and $W(\tau)=U^{\dagger}(\tau)W(0)U(\tau)$ is a time-evolved observable at $t=\tau$ supported on a larger number of sites, in the Heisenberg picture.
Then, the OTOC is defined by $C_{\beta}(\tau):=-\expt{[W(\tau),V(0)]^{2}}_{\rho}$ at temperature $T=\beta^{-1}$ \cite{Maldacena16}.
The initial state $\rho$ is usually considered to be the maximally mixed state, i.e. $\rho\propto\openone$, in which case $\expt{\bullet}_{\rho}$ stands for the thermal average.

Using these notations, we move on to outline the results for the OTOC.
\begin{theorem}\label{th:OTOC}
Suppose $W$ is the self-adjoint and unitary operator while $V$ is the self-adjoint operator.
Then, the OTOC is represented by
\eq{
\label{eq:specific_dist=OTOC}
C_{\beta}(\tau)&=\lim_{\theta\ra0}\frac{\delta^{2}(\calL_{\rho,V,\theta,\calD_{W}},\Omega_{1/2,\pm},\calR_{V,\bfS})}{\theta^{2}}\nonumber\\
&=\eta^{2}(\rho,V,\calD_{W},\calR_{V,\bfS}),
}
where $\calL_{\rho,V,\theta,\calD_{W}}:=(\calD_{W}\otimes\openone_{\bfQ})\circ\calU_{V,\theta}\circ\calA_{\rho}$ is the loss process with $\calD_{W}(\bullet):=W(\tau)(\bullet)W^{\dagger}(\tau)$ and $\calR_{V,\bfS}:=\calJ_{\bfS}\circ\calU^{\dagger}_{V,\theta}$ is the recovery process.
\end{theorem}
The situation considered above is depicted in Fig.~\ref{fig:Quantum_comb_otoc}; and Theorem~\ref{th:OTOC} can also be extended to the case where $W$ is the self-adjoint operator \cite{Note1}.
It is noteworthy that this IEP works as an experimental evaluation method for the OTOC.
To overcome some difficulties concerning an implementation, several approaches for measuring the OTOC are proposed: rewinding time \cite{Swingle16,Swingle18}, interferometer \cite{Zhu16,Yao16}, weak-measurement \cite{Halpern17,Halpern18,Dressel18}, two-point measurement \cite{Campisi17}, R\'{e}nyi entropy \cite{Fan17}, multiple quantum coherence intensity \cite{Wei18}, quantum teleportation \cite{Yoshida19}, randomized measurement \cite{Vermersch19}, thermofield double state \cite{Hurtubise20}, classical shadows \cite{Garcia21}, and so forth.
Most of these methods require a number of measurements at each time point during the time evolution in the whole system, with several prepared measuring devices.
In contrast, our approach only requires a single measuring device for measuring on $\bfQ$ at the end of the process.
In this regard, our method has the advantages of measuring at one point, simplifying the setup and being easy to implement.
As a demonstration, we show experimental results based on our method and other methods for comparison in the companion paper \cite{Emori25}.
As a natural consequence, we can come by a universal lower bound as $\eta^{2}(\rho,V,\calD_{W})\le C_{\beta}(\tau)$ and the WAY theorem for the OTOC.
The proof of Theorem~\ref{th:OTOC} and its generalized version are detailed in the Supplemental Material \cite{Note1}.

\ti{\textbf{Outlook.\md}}Basically, our formulation can be used for various physical processes and some sort of quantities, e.g. entropies \cite{Buscemi14}, in their frameworks as irreversibility beyond error, disturbance, and OTOC.
Specifically, by leveraging the knowledge of irreversibility accumulated in physics, the recent rapid developments in stochastic thermodynamics and quantum information theory can be the scope of applications of our formulation.
It provides a fertile ground for and opens a lot of new possibilities for applications; we intend to explore these aspects in future work with promising prospects for further advancements.

\begin{acknowledgments}
\ti{\textbf{Acknowledgments.\md}}The authors would like to thank Akihisa Tomita for valuable discussions and Satoshi Nakajima and Yui Kuramochi for helpful comments.
HE was supported by JST SPRING, Grant Number JPMJSP2119, and RIKEN Junior Research Associate Program.
HT was supported by JSPS Grants-in-Aid for Scientific Research 
No. JP25K00924, and MEXT KAKENHI Grant-in-Aid for Transformative
Research Areas B ``Quantum Energy Innovation'' Grant Numbers 24H00830 and 24H00831, JST MOONSHOT No. JPMJMS2061, and JST FOREST No. JPMJFR2365.
\end{acknowledgments}

\bibliography{ref}

\begin{figure}[tb]
\centering
\includegraphics[width=\linewidth]{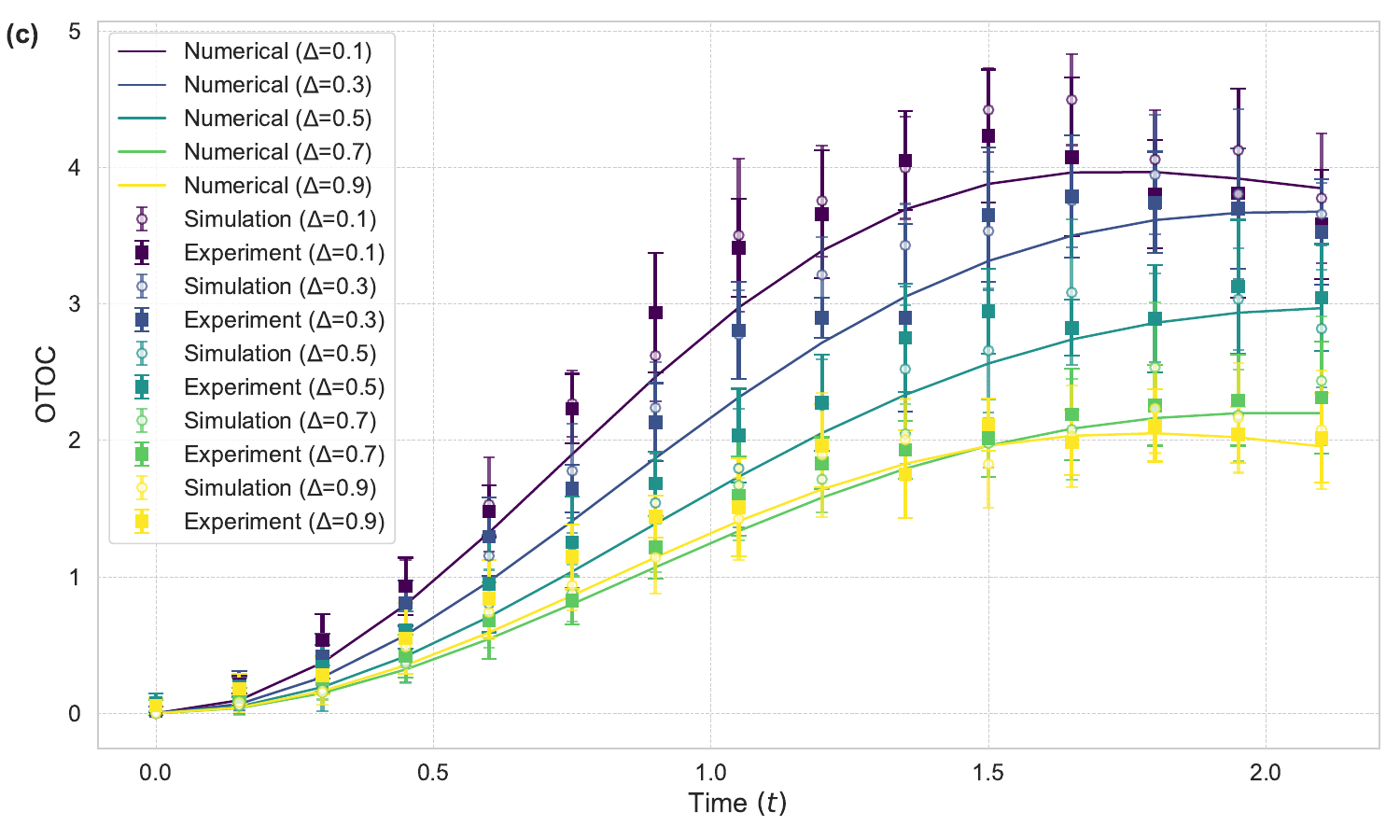}
\caption{
\label{fig:res_otoc}
The experimental result of the measurement of the OTOC using our method.
The results present a comparison among the ideal values (solid lines) from matrix calculations, the noiseless values (circle dots) from \texttt{aer-simulator}, and the experimental results (square dots) from \texttt{reimei}.
The error bars represent the standard deviation of the measured values.
The results show good agreement with theory, particularly at early and late times, validating our method.
}
\end{figure}

\section{End Matter}
We report on the experimental evaluation of the OTOC using a trapped-ion quantum computer---the \texttt{reimei} of Quantinuum \cite{Quantinuum}.
The full comparative study, including the rewinding time method (RTM) \cite{Swingle16,Swingle18} and the weak-measurement method (WMM) \cite{Halpern17,Halpern18,Dressel18}, is presented in a companion paper \cite{Emori25}.
Here, we detail the result, focusing on our proposed method, which is referred to irreversibility-susceptibility method (ISM).

\textit{\textbf{State preparation.---}}Measuring the OTOC in a thermal environment requires preparing the Gibbs state with a given Hamiltonian $H$ and an inverse temperature $\beta = 1/(k_{B}T)$, defined as
\eq{
\rho_{\text{Gibbs}}=\frac{e^{-\beta H}}{Z(\beta,H)},
}
where $Z(\beta,H)=\Tr[\exp(-\beta H)]$ is a partition function.
It is a mixed state and thus not unitarily preparable on a closed system.
To overcome this difficulty, we have employed a variational quantum algorithm (VQA) \cite{Consiglio24} that prepares the thermofield double (TFD) state on an enlarged system $\mathbf{S} + \mathbf{A}$, where $\mathbf{S}$ is the $n$-qubit target system and $\mathbf{A}$ is an $n$-qubit ancilla system.
The Gibbs state $\rho_{\text{Gibbs}}$ is the unique state that minimizes the Helmholtz free energy $F(\rho)=\Tr(H\rho)-\beta^{-1}S(\rho)$ such that
\eq{
\rho_{\text{Gibbs}}=\arg\min_{\rho}F(\rho),
}
where $S(\rho)$ is the von Neumann entropy $S(\rho)=-\Tr(\rho\log\rho)$.

The key insight is that for the TFD state, the entropy of $S(\rho_{\mathbf{S}})$ equals the entropy of $S(\rho_{\mathbf{A}})$.
Since $\rho_{\mathbf{A}}$ is diagonal in the computational basis, its entropy can be efficiently calculated classically from probabilities of computational basis measurement on $\mathbf{A}$.
Therefore, by defining the cost function as the free energy of $\mathbf{S}$,
we just need to minimize the cost function, and it yields parameters $\bm{\theta}^{*}$ and $\bm{\phi}^{*}$ that prepare the desired approximation of $\rho_{\text{Gibbs}}$ on $\mathbf{S}$.

\textit{\textbf{Hamiltonian and observables.---}}The system dynamics are generated by the one-dimensional Heisenberg XXZ model with a transverse magnetic field
\eq{
H=-\frac{1}{4}\sum^{n-1}_{i=1}(\sigma^{x}_{i}\sigma^{x}_{i+1}+\sigma^{y}_{i}\sigma^{y}_{i+1}+\Delta\sigma^{z}_{i}\sigma^{z}_{i+1})-h\sum^{n}_{i=1}\sigma^{z}_{i},
}
where $\sigma^{\bullet}_{i}$ for $\bullet=x,y,z$ are the Pauli matrices acting on the $i$-th qubit, $\Delta$ is the anisotropy parameter, and $h$ is the strength of the transverse magnetic field.
In our experiment, we set the magnetic field strength to $h=(1-\Delta)/2$ and the number of qubits to $n=4$.
We have scanned $\Delta \in \{0.1, 0.3, 0.5, 0.7, 0.9\}$ to explore regimes spanning from near-integrable to chaotic.
The operators of the OTOC are chosen as the local Pauli-$x$ operator on the first qubit: $W(0)=V(0)=\sigma^{x}_{1} \equiv X\otimes\openone\otimes\openone\otimes\openone$.
The time evolution is measured up to $t=2.1$, discretized into $15$ time steps.
Each data point is averaged over $10$ experimental iterations, with $1000$ measurement shots per circuit execution.
The strength of weak interactions is set to $\theta=0.4$ radians.

\textit{\textbf{Experimental results of measuring the OTOC.---}}
The results of measuring the OTOC via the ISM are summarized in Fig.~\ref{fig:res_otoc}.
The experimental data (square dots) align well with the ideal theoretical curves (solid lines) and noiseless simulations (circle dots) for all tested values of $\Delta$ on average
A detailed comparison with existing approaches is presented in the companion paper \cite{Emori25}.
The main characteristics of the different methods can be summarized as follows.
RTM and WMM tend to show large deviations between the theoretically predicted values and those obtained experimentally, typically in the late-time regime.
By contrast, our method does not exhibit such large systematic deviations and reproduces the experimentally observed values well across the entire time range.
On the other hand, our method tends to yield a larger standard deviation than the existing approaches.
This is a consequence of the fact that our protocol relies on the weak interactions.
For this reason, the existing methods and our method should be regarded as complementary.
In particular, since the standard deviation can be reduced by increasing the number of shots, our method becomes a more accurate approach for predicting the OTOC in regimes where sufficient sampling is available, as it reproduces the correct values over all times.
In summary, our data successfully demonstrate the practical feasibility of experimentally evaluating the OTOC on current-generation quantum hardware.

\clearpage
\onecolumngrid

\foreach \x in {1,...,25} {
  \includepdf[pages=\x]{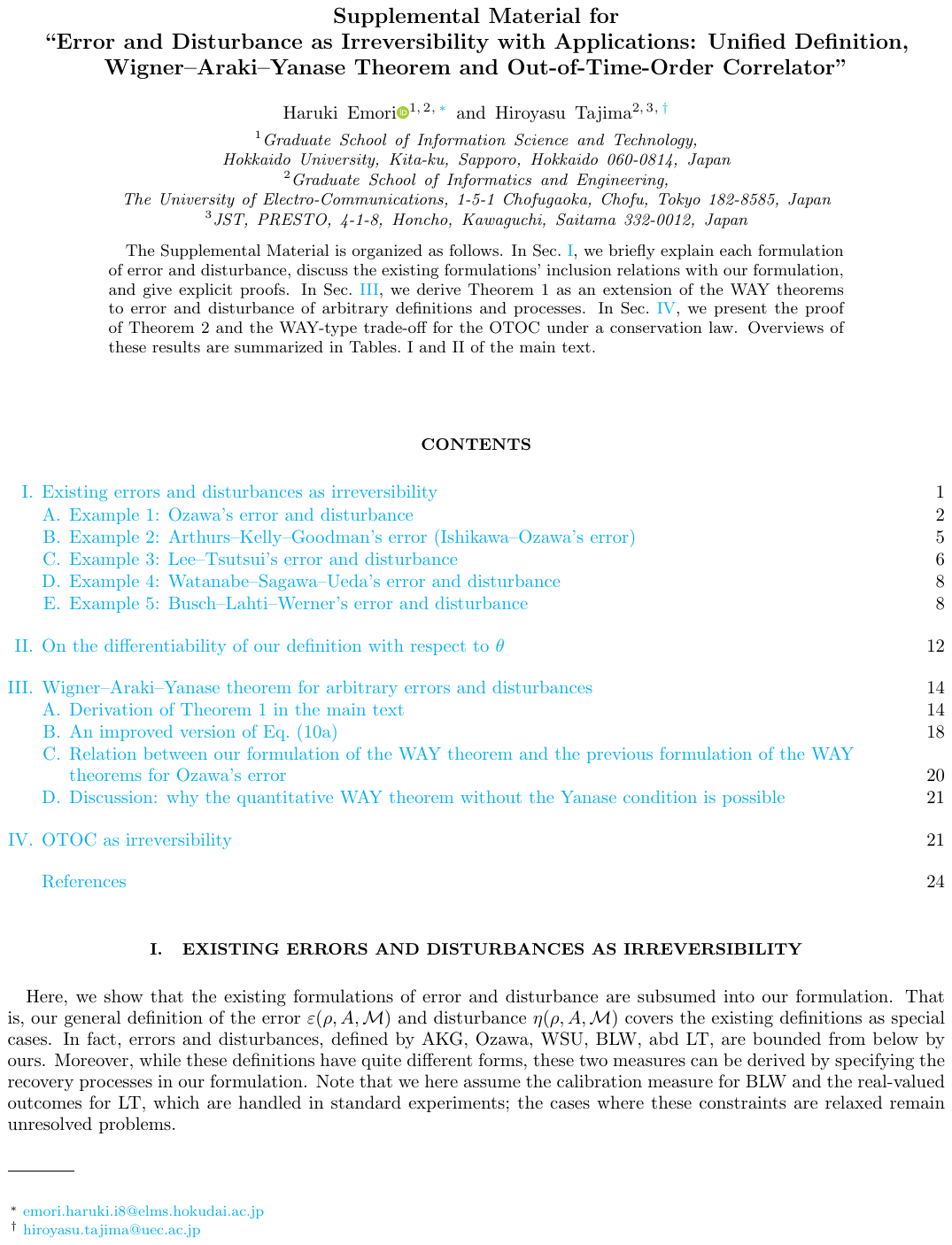}
}

\end{document}